\documentclass
[
amssymb,prb,twocolumn,floatfix,tightenlines,superscriptaddress]{revtex4-1}
\pdfoutput=1
\usepackage{graphics}
\usepackage{graphicx,color}
\usepackage{subfigure}
\usepackage{float}
\usepackage{graphicx}
\usepackage{amsmath}
\usepackage{amssymb}
\usepackage{dsfont} 
\usepackage{hyphenat}
\usepackage[percent]{overpic}
\usepackage{xcolor}
\usepackage{appendix}
\usepackage{textcomp}

\newcommand{\ket}[1]{\left | #1 \right \rangle}

\newcommand{\blueit}[1] {\textcolor{black}{#1}}
\newcommand{\redit}[1] {\textcolor{black}{#1}}

\begin{document}

\title{Testing foundations of quantum mechanics with photons}
\author{Peter Shadbolt, Jonathan C. F. Mathews, Anthony Laing, and Jeremy L. O'Brien}
\affiliation{Centre for Quantum Photonics, H. H. Wills Physics Laboratory \& Department of Electrical and Electronic Engineering, University of Bristol, Merchant Venturers Building, Woodland Road, Bristol, BS8 1UB, UK}
\date{\today}

\begin{abstract}
The foundational ideas of quantum mechanics continue to give rise to counterintuitive theories and physical effects that are in conflict with a classical description of Nature.  Experiments with light at the single photon level have historically been at the forefront of tests of fundamental quantum theory and new developments in photonics engineering continue to enable new experiments.
Here we review recent photonic experiments to test two foundational themes in quantum mechanics:
wave-particle duality, central to recent complementarity and delayed-choice experiments; and 
Bell nonlocality where recent theoretical and technological advances have allowed all controversial loopholes to be separately addressed in different photonics experiments.
\end{abstract}

\maketitle
\noindent
Light has featured in tests of foundational physics during times that witnessed major advancements in our understanding of nature.
Newton's investigation of the nature of light, using prisms to reveal the visible spectrum, is iconic of the Scientific Revolution of the 16th and 17th centuries. 
Early tests of Einstein's general relativity involved observations of starlight passing close to the sun during a solar eclipse.  Scientific advances have led to more convincing (and striking!) observations of relativistic effects with images from the Hubble space telescope revealing the gravitational lensing of galactic light. And we now know how the electromagnetic spectrum extends beyond the visible range and is quantised into single photons.  Since the primary detection apparatus in early experimental physics were the physicists themselves, light made a natural observable.  As our understanding of quantum photonics deepened, the utility of photons in tests of foundational concepts in physics became more evident.  Photons are robust to environmental noise, have low decoherence properties, and are easily manipulated and detected.  The first half of this review discusses tests of wave-particle duality with one photon, while the second half looks at experimental tests of nonlocality with two or more photons.\\

\noindent
\textbf{1. Wave-particle duality}\\
\nopagebreak
The double-slit experiment has famously been said to contain the entire mystery of quantum mechanics. It provides a concise demonstration of the fact that single quanta are neither waves nor particles, and that in general they are neither in one single place, nor in two places at once.  

The experiment begins with a source of single quanta. Here we will consider only photons, but qualitatively identical results have been observed in a wide variety of quantum systems including electrons \cite{Tonomura1989, Bach2013, Jonsson1974}, atoms \cite{Carnal1991}, and even large molecules such as $C_{60}$ \cite{Arndt1999}.
Single photons are sent towards a mask into which two slits have been cut. On the far side of the mask, the spatial distribution of single-photon detection events is measured by a sensitive detector. For each photon, the detector registers a ``click'' at position $x$.  Simultaneous detection of two clicks never occurs, and the photon initially appears to travel and arrive as a discrete particle. However, having detected many photons, the observed probability distribution $p(x)$ can only be explained by wave interference, due to components of the photon which travel through \emph{both} slits simultaneously. Confoundingly, when detectors are placed directly inside the two slits, the photon is only ever detected at one slit or the other --- never at both.

Where was the photon when it travelled through the mask? If it passed through one slit and not the other, wave interference effects would not be observed. If it passed through both slits at once, it should be possible to detect it at both simultaneously; this never occurs. If it passed through neither slit, we should not detect it at all --- but we do. In this way, the double-slit experiment reveals the inadequacy of classical language when describing quantum systems.

In 1909, Geoffrey Taylor used a sewing needle to split a beam of light into two paths, and observed interference fringes in the resulting pattern of light and shadow \cite{Taylor1909}.  He used an incandescent source of ``feeble light'' with roughly the intensity of a candle held at a distance of one mile.  
Since then, single photons have played a pivotal role in tests of wave-particle duality. 
This is largely due to the ease with which photons can be generated, manipulated and measured, as well as their resilience to noise, and associated room temperature/pressure operation. 
Many of these experiments are based on a very natural question: what do we know, and how much can we measure, of the state of the photon as it passes through the slits?

The light source used by Taylor was \emph{thermal} --- it did not generate photons one-by-one --- and his experiment consequently admits a classical model.  The fact that true single photons are not detected at both slits simultaneously (\emph{antibunching}) was confirmed experimentally by Clauser \cite{Clauser1974}, who used a more sophisticated light source, based on atomic cascades in mercury atoms.  A similar source was used by Grangier et al. \cite{Grangier1986}, who observed both antibunching and wave interference effects analogous to those of the double-slit experiment. 

In the quantum-mechanical description, detection of the photon at one slit ``collapses'' the single-photon wavefunction and precludes detection at the other slit. Collapse is instantaneous, even when the slits are very far apart, and it was emphasised by Einstein at the Solvay conference \cite{Jammer1974} that the effect is thus seemingly \emph{nonlocal}. A recent experiment by Guerreiro et al. \cite{Guerreiro2012} tested Einstein's thought experiment for the first time, using space-like separated (causally independent) detectors. 

Wavefunction collapse belongs to the Copenhagen interpretation of quantum mechanics, and \redit{encompasses} Niels Bohr's principle of \emph{complementarity}. Bohr maintained that in order to observe complementary properties of a quantum system, an experimentalist must necessarily employ mutually incompatible arrangements of the measurement apparatus. In the context of the double slit, this means that any experiment which fully reveals the wave-like properties of the photon must obscure its particle-like character, and vice-versa.  

Bohr's principle has only very recently been formalized in \emph{universal complementarity relations}, such as those due to Ozawa and Hall \cite{Hall2004, Ozawa2003, Erhart2012}.
These relations formalize the notion that although the inaccuracy in either of two complementary observables can individually be made arbitrarily small, one cannot simultaneously measure both to an arbitrary degree of accuracy. This is distinguished from the Heisenberg uncertainty principle, where measurements are sequential rather than simultaneous.
Very recently, Weston et al. \cite{Weston2013}, used a spontaneous parametric down-conversion source together with a linear-optical circuit to experimentally test these new relations. Making use of entanglement generated by the photon source, the authors were able to test complementarity under conditions in which previously discovered, non-universal complementarity relations fail.

Upon first encountering the double slit experiment, it is natural to wonder about the \emph{trajectory} of the photon during its path from source to detector. Complementarity implies that it is not possible to simultaneously measure the position of the photon without irrevocably disturbing its momentum, and indeed a na\"ive experiment hoping to track the route of the photon by measurement of its position will destroy all wave-like effects. However, Kocsis et al. recently demonstrated~\cite{Wiseman2007, Kocsis2011} that \emph{quantum weak measurement}~\cite{Aharonov1988} can be used to approximately reconstruct the average trajectory of ensembles of photons as they undergo double-slit interference. Weak measurement allows approximate information to be obtained on a particular observable without appreciably disturbing ``strong'' measurement outcomes on a complementary variable. 
The authors sent single photons from a GaAs quantum dot through a double-slit interferometer, in which a piece of birefringent calcite imposes a weak polarization rotation depending on the angle of incidence --- and thus the momentum --- of the photon. Then, by simultaneous detection of the lateral position and polarization using a high-resolution CCD camera, weak measurement of the photon's momentum was accomplished at the same time as strong measurement of position. The trajectories measured in this experiment hold particular significance in the de Broglie-Bohm interpretation of quantum mechanics, where they are literally interpreted as the path taken by a single particle-like photon.

\begin{figure}[h]
\centering
\includegraphics[width=\columnwidth]{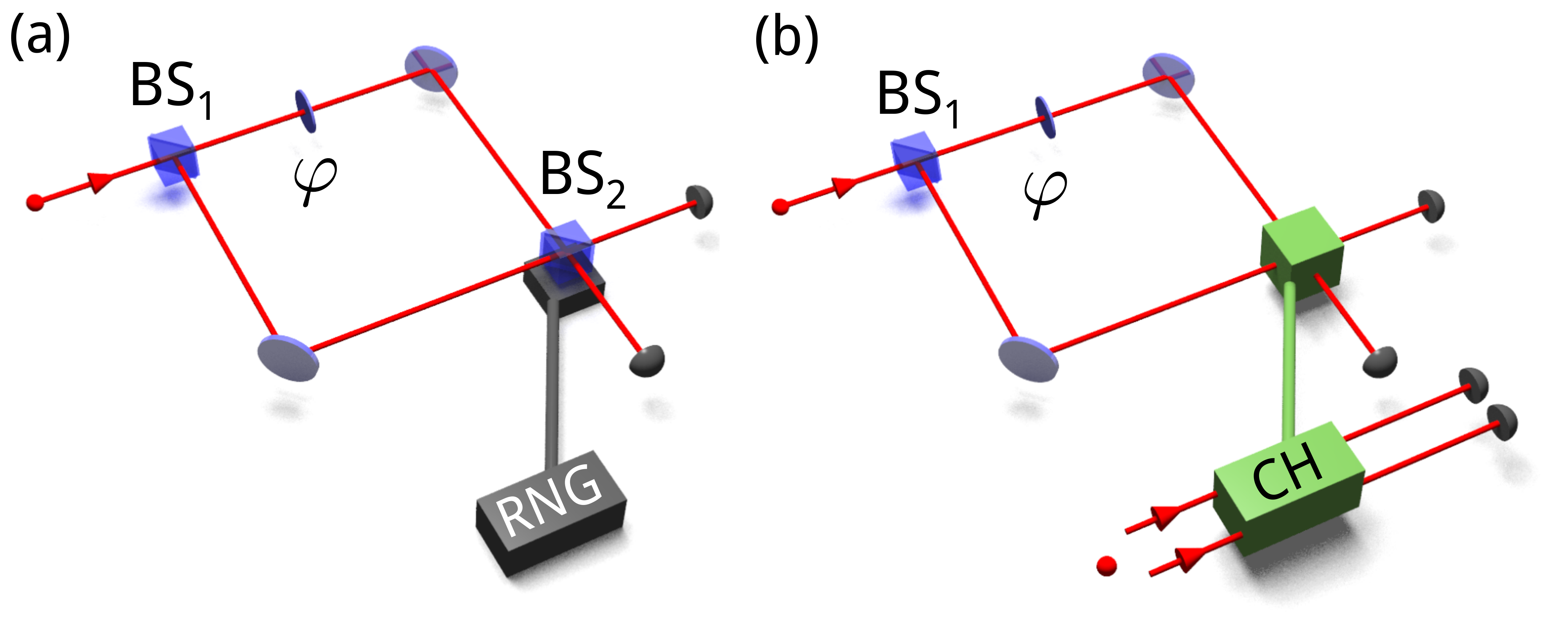}
\caption{Delayed choice experimental set-ups.
(a) Wheeler's delayed choice experiment. A photon is sent into a Mach-Zehnder interferometer. Upon arrival at the first beamsplitter $BS_1$, it is split into a superposition across both paths. A random number generator (RNG) then toggles a fast optical switch, closing or opening the interferometer by insertion or removal of $BS_2$, leading to wave-like or particle-like measurement of the photon respectively. Two detectors reveal wave-like behaviour in the event that the interferometer is closed, otherwise particle-like statistics are seen.
(b) Quantum delayed choice. The optical switch is replaced by a quantum-controlled beamsplitter: a controlled-Hadamard gate. An ancilla photon controls this gate: ancilla states $\ket{0}$ and $\ket{1}$ lead to presence and absence of $BS_2$ respectively. By preparing the ancilla in a superposition state $BS_2$ can be placed into a superposition of present and absent, leading to a superposition of wave-like and particle-like measurement.}
\label{wpd}
\end{figure}

John Wheeler's famous \emph{delayed-choice} thought experiment~\cite{Wheeler1978, Wheeler1984} also addresses the question of the position or trajectory of the photon in a two-path setup. Considering the double slit experiment, one might attempt to side-step the uncomfortable implications of wave-particle duality by means of a pseudo-classical explanation in which the photon \emph{decides in advance} to behave as a particle or wave, depending on the choice of measurement setup. If the photon notices that a particle-like measurement is planned, it dispenses with all wave-like properties and passes through one slit at random, and vice-versa.  Wheeler proposed an elegant test of this comforting (if pathological) model, in which the decision to measure wavelike or particle-like behaviour is delayed until \emph{after} the photon has passed the slits, but before it reaches the measuring apparatus.  Delayed-choice experiments have been performed in a variety of physical systems~\cite{Alley1987, Hellmuth1987,Lawson-Daku1996, Kim2000}, all of which confirm the quantum predictions and refute the notion that the photon decides in advance to behave as a particle or a wave. 

Of particular significance is a recent result~\cite{Jacques2007} of Jacques et al., in which relativistic space-like separation between the random choice of measurement setting and slits was achieved for the first time. This ensures that there can be no causal link between the free choice of measurement setting and the behaviour of the photon at the slits.  Here, a nitrogen vacancy colour centre in diamond was used as the source of single photons, ensuring extremely close approximation to the single-photon Fock state $\ket{1}$. An electro-optic modulator, driven by a quantum random number generator at 4.2MHz, was used to implement the choice of measurement setting. A similar experimental setup was more recently employed by the same group \cite{Jacques2008} to refute the controversial claims due to Afshar et al. \cite{Afshar2005, Afshar2007} that Bohr's complementarity principle could be violated in a subtle variation on the double slit experiment.

In delayed-choice experiments, the choice of measurement setting is generally implemented using a classical optical switch, driven by a random number generator, which rapidly inserts or removes an optical beam-splitter in the path of the photon. If the beam-splitter is present, which-way information is erased and full-contrast wavelike interference is observed. If the beamsplitter is instead absent, each detection event yields full which-way information but no interference is seen. A recent proposal by Ionicioiu and Terno \cite{Ionicioiu2011c} suggested that the classical random bit might be replaced by an ancilla qubit, and the classical controlled-beamsplitter by a quantum controlled-beamsplitter, or controlled-Hadamard ($CH$) gate. By preparing the ancilla qubit in the superposition state $\cos(\alpha)\ket{0} + \sin(\alpha)\ket{1}$, the beamsplitter is effectively placed into a coherent superposition of \redit{being} present and \redit{being} absent. One can then continuously tune between particle-like and wavelike measurement settings, in close analogy with the weak measurement technique of Ref.~\onlinecite{Kocsis2011}. This idea was quickly implemented by a number of groups \cite{Kaiser2012, Peruzzo2012, Singha2012}, two of which used photon pairs generated by SPDC. The result of Peruzzo \textit{et al.}\cite{Peruzzo2012} exploits recent developments in integrated quantum photonics \cite{Politi2009}, with Wheeler's interferometer and the $CH$ gate both implemented on-chip \cite{Shadbolt2012}. In this experiment, entanglement generated by the $CH$ gate allows for device-independent refutation of hidden variable models in which the photon decides in advance to behave as a particle or a wave, by violation of the Bell-CHSH inequality \cite{Clauser1969}.

In the scenario of Ionicioiu and Terno, which-way information is carried by the ancillary particle. This possibility was previously emphasized by Scully and Dr\"uhl \cite{Scully1982}, who pointed out that the choice of measurement basis for the entangled ancilla determines the contrast of wave interference observed, and that this choice can be made even \emph{after} the system photon has been detected. Only a subset of allowed measurement settings completely and irrevocably erase all which-way information, resulting in high-contrast interference fringes. In 2000, Kim et al. \cite{Kim2000} implemented this so-called \emph{delayed-choice quantum eraser} using single photons generated by SPDC.  More recently, Ma et al. \cite{Ma2013} demonstrated a quantum eraser using entangled photon pairs. 
The authors went to great lengths to rule out models in which the circumstances of the ancilla are communicated to the system photon through a local, causal mechanism. The system and ancilla photons were sent to separate islands, 144km apart, such that the choice of measurement setting, system photon and interferometer, and measurement of the ancilla photon, were all mutually space-like separated --- and therefore causally disconnected.

In all of these variants on the double-slit experiment, we see a fundamental trade-off between the information which can simultaneously be obtained on particular properties of quantum systems, as well as a behavioural dependence on the choice of measurement setting. These effects seem contrary to what is known as \emph{noncontextual realism}, the (rather natural) assumption that the observable properties of objects are well-defined independent of measurement.  Kochen and Specker (KS) \cite{Kochen1967} proved that there exist sets of quantum-mechanical observables, to which a unique set of values provably cannot be consistently and simultaneously assigned --- rendering noncontextuality untenable. No direct experimental implementation has yet been reported, and indeed it is unlikely that a meaningful direct implementation of KS is possible \cite{Cabello1998, Meyer1999}. However, a number of theoretical works \cite{Greenberger1990, Simon2000, Cabello2001, Amselem2009} have shown that noncontextual realism can be revoked under much less demanding conditions, and many of these theories have since been tested using single photons.  An early result by Michler et al. \cite{Michler2000} mimicked three-particle GHZ correlations, using entangled photon pairs generated by SPDC.  Huang et al. \cite{Huang2003} tested the ``all-or-nothing'' KS-like theory of Simon \cite{Simon2000}, encoding two qubits in the path and polarization degrees of freedom of one heralded single photon. More recently, Lapkiewicz et al. reported \cite{Lapkiewicz2011} an experiment using a single photonic qutrit, encoded in path and polarization using calcite beam displacers, to implement the theoretical proposal of Klyachko et al. \cite{Klyachko2008}. This result is notable as it reinforces the strong incompatibility between the quantum and classical pictures of physics with only a single quantum particle and in the absence of multi-particle entanglement.

A fundamental tenet of quantum mechanics is the \emph{Born rule}, which states that given a system with wavefunction $\psi(\vec{r}, t)$, the probability that it is detected in the volume element $d^3r$ at time $t$ is given by 
\begin{equation}
p(\vec{r}, t) = |\psi(\vec{r}, t)|^2 d^3 r .
\label{eqn:born}
\end{equation}
It can easily be shown that since this expression depends only on the \emph{square} of the wavefunction, probabilities generated by multi-particle wavefunctions can always be written in terms of interference between pairs; three-body interference terms never appear in the expansion of (\ref{eqn:born}). Indeed, it has been shown that almost all nonlinear models of quantum mechanics which permit three-body interference have extreme and highly unlikely consequences, such as cloning of quantum states and polynomial-time quantum algorithms for NP-complete computational tasks \cite{Abrams1998}. A recent experiment by Sinha et al. \cite{Sinha2010} went in search of such effects using a triple-slit variation on the double-slit experiment. Using a lithographically fabricated triple slit, a coherent laser source, and heralded single photons from SPDC, the authors gathered strong evidence against the existence of higher-order corrections to the Born rule.\\

\noindent\textbf{2. Nonlocality}\\
{
\emph{Locality} is the concept that the behaviour of space-like separated objects depends only on events in their respective light-cones.
Confoundingly, entangled particles exhibit correlations that defy this understanding.
Many attempts have been made to explain these correlations in terms of \emph{local hidden variable} models (LHVs) which attempt to capture our everyday experience of the universe.
LHVs associated to each particle can be imagined as having been determined from some {earlier local interaction}. This aligns with an intuitive {\emph{local} and \emph{realistic}} view of a universe that is causally connected by locality.
In 1964, John Bell described an experimentally tenable scenario in which quantum mechanics predicts outcomes that are incompatible with all possible LHV models~\cite{Bell1964c}---provided the experiments are rigorously performed.
} 
The {platform} of {entangled} single photons 
is the only platform {to have} addressed all the {known key} requirements {of} a quantum theory of nonlocality, albeit in separate experiments.
Here we review a selection of recent developments using entangled photons to test quantum nonlocality and explore its properties. {For an exhaustive review of the subject's history} we point the reader towards more in-depth reviews on multi-photon entanglement~\cite{pa-rmp-84-777} and theoretical developments~\cite{Brunner2013}.

\begin{figure}[b!]
\centering
\includegraphics[width=\columnwidth]{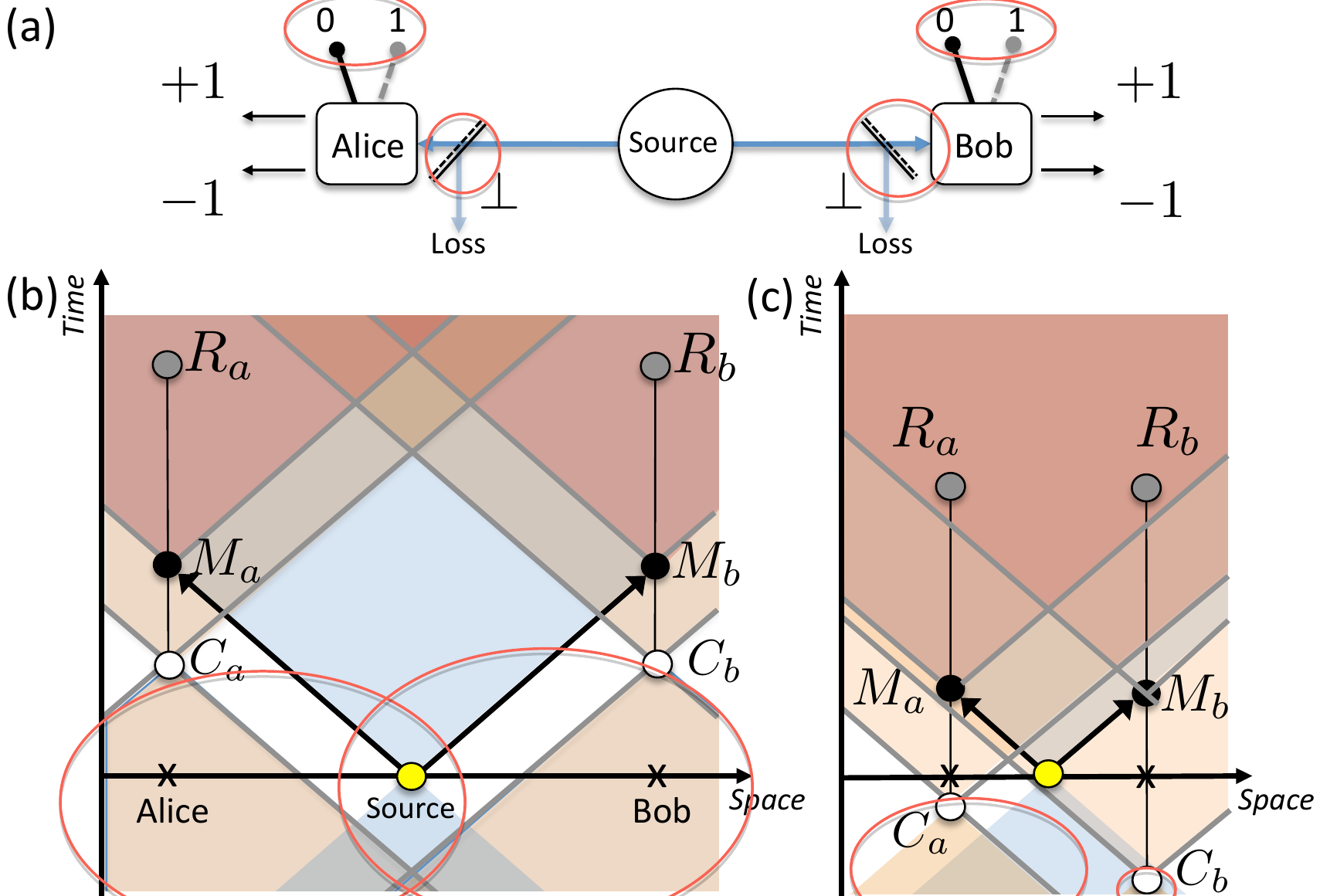}
\caption{{\textbf{A nonlocality experiment and associated loopholes.} (a) The detection loophole can be opened by optical loss if there is a sufficiently high proportion of non-conclusive outcomes ``$\perp$''. (b) A space-like separation prohibits signalling between the various events occurring for each observer and closes the locality loophole. E.g. Alice's measurement $M_a$ and results $R_a$ are outside of the light-cone of influence from Bob's measurement choice $C_b$. Furthermore since $C_a$ and $C_b$ are causally disconnected from detection events and the source, Alice and Bob are free to choose their measurement settings without influence. (c) If the observers are not space-like separated, it is possible for signalling to occur between events. In this example, $M_a$ and $M_b$ can respectively influence $R_b$ and $R_a$, and $C_b$ can influence both $M_a$ and $R_b$.
}}
\label{LoopholesFig}
\end{figure}

Since the seminal experiments of Freedmand and Clauser \cite{fr-prl-28-938} in 1972, and Aspect et al. \cite{Aspect1982} in 1982,
nonlocality experiments have typically comprised a source of entangled photon pairs that are shared between observers Alice and Bob who independently perform measurements and subsequently compare their results (Fig.~\ref{LoopholesFig} (a)). The measurements have two possible settings, $0$ or $1$, and have only two possible outcomes $a, b \in \{-1,+1 \}$, \blueit{often assigned to the polarisation of the photons}.
The Clauser-Horne-Shimony-Holt (CHSH) \cite{Clauser1969} version of Bell's inequality sets an upper bound on the strength of correlations allowed by LHV models using the \blueit{a sum of} expected values for $ab$, for each possible \blueit{combination of} measurement outcome:
\begin{eqnarray}
\blueit{S = |}\left\langle a_0 b_0 \right\rangle + \left\langle a_0 b_1 \right\rangle + \left\langle a_1 b_0 \right\rangle - \left\langle a_1 b_1 \right\rangle \blueit{|}\leq 2.
\label{CHSH}
\end{eqnarray}
Quantum mechanics predicts that this bound can be \blueit{experimentally} violated, demonstrating the inadequacy of LHV models. If the two particles are entangled---for example in the singlet state $\ket{\psi^-} =\frac{1}{\sqrt{2}}\left(\ket {0 1} - \ket{1 0}\right)$---\blueit{then a choice of measurement settings $\sigma_{z}$ \& $\sigma_{x} $ for Alice, and $\sigma_{z} + \sigma_{x}$ \& $\sigma_{z} - \sigma_{x}$ for Bob, leads to a violation of (\ref{CHSH}) with $S = 2 \sqrt{2}$.}

\blueit{The implications of rigorously violating this inequality have a profound effect on our intuition of how the universe works, for it suggests that the two particles are instantaneously communicating with one another, even though they are far apart.  Although the randomness of outcomes to measurements mean that no communication can occur between Alice and Bob, these nonlocal effects seem to be in contradiction with the spirit, if not the letter, of special relativity.  These far reaching implications have motivated particular scrutiny on the possible ways in which nature might somehow \emph{fake} nonlocality, with focus mainly falling upon experimental limitations. An apparent experimental violation $S >2$ could be attributed to assumptions exploited by LHV models and known as \emph{loopholes}}, the more famous of which are the \textit{Locality}, \textit{Detection} and \textit{Freedom of Choice} loopholes (Fig.~\ref{LoopholesFig}). \blueit{A completely unambiguous experimental demonstration of} Bell nonlocality requires the simultaneous obstruction of every possible loophole. While this milestone is yet to be reached in experimental physics, photons have been used to address each of these loopholes \blueit{individually}.

\emph{The Detection Loophole---}Optical tests of nonlocality have suffered from low detection efficiency. With an experiment\blueit{al} efficiency of \mbox{$\eta<100\%$} there exist, \blueit{in addition to ``$+1$'' and ``$-1$'', non-conclusive measurement outcomes ``$\perp$''} that represent the failure to detect an emitted photon. ``$\perp$'' can be ignored by \blueit{including only} measurements that register photon detection. But this \blueit{relies on the assumption of a} \textit{fair sampling}, since otherwise local models may skew the detection statistics of ``$+1$'' and ``$-1$'' to falsify violation of (eq.~\ref{CHSH}). This has been illustrated experimentally through the use of side channels to intentionally falsify signatures of nonlocality in experimental setups that are otherwise considered as standard Bell-inequality experiments \cite{ta-pra-80-030101, ge-prl-107-170404,po-njp-13-063031}.

When including ``$\perp$'' outcomes, violation of CHSH (eq.~\ref{CHSH}) only occurs when experiment efficiencies are beyond the threshold of $\eta > 82.8\%$. Remarkably, Eberhard discovered that lowering the amount of entanglement by controlling the $r$ parameter in \blueit{$(r\ket{01} - \ket{10}) / \sqrt( 1 + r^2)$} reduces the threshold efficiency to $\eta > 66.7\%$ in testing nonlocality \cite{eb-Pra-47-747}. Denoting $n_{k,l}(a_i b_j)$ as the number of photon pairs with outcome $k \in \{ +1, -1, \perp \}$ and  $l \in \{ +1, -1, \perp \}$ when using measurement settings $i\in\{0,1\}$ on one particle and $j\in\{0,1\}$ on the other, Eberhard's inequality (which holds for LHV) is written as
\begin{eqnarray}
J\!& = & \! n_{+1,+1}(a_1,b_1)-n_{+1,+1}(a_0,b_0)+n_{+1,-1}(a_0,b_1)\label{Eberhard}\\
&&n_{+1,\perp}(a_0,b_1)+n_{-1,+1}(a_1,b_0)+n_{\perp,+1}(a_1,b_0)\geq 0\nonumber
\end{eqnarray}
Notably, each observer need\blueit{s} only one detector, since the decrease in efficiency of detectors responsible for ``$-1$'' outcomes causes \blueit{outcomes nominally} ``$-1$'' to be included as ``$\perp$'', mapping $n_{+1,-1}(a_0, b_1)$ to $n_{+1,\perp}(a_0, b_1)$ and $n_{-1,+1}(a_1, b_0)$ to $n_{\perp,+1}(a_1, b_0)$ \cite{gi-nat-497-227}. Furthermore, testing nonlocality with (eq.~\ref{Eberhard}) is robust to the \blueit{Poissonian} nature of photon-counting measurements on SPDC sources and removal of the vacuum via post-selection is not required.

\begin{figure}[b!]
\centering
\includegraphics[width=\columnwidth]{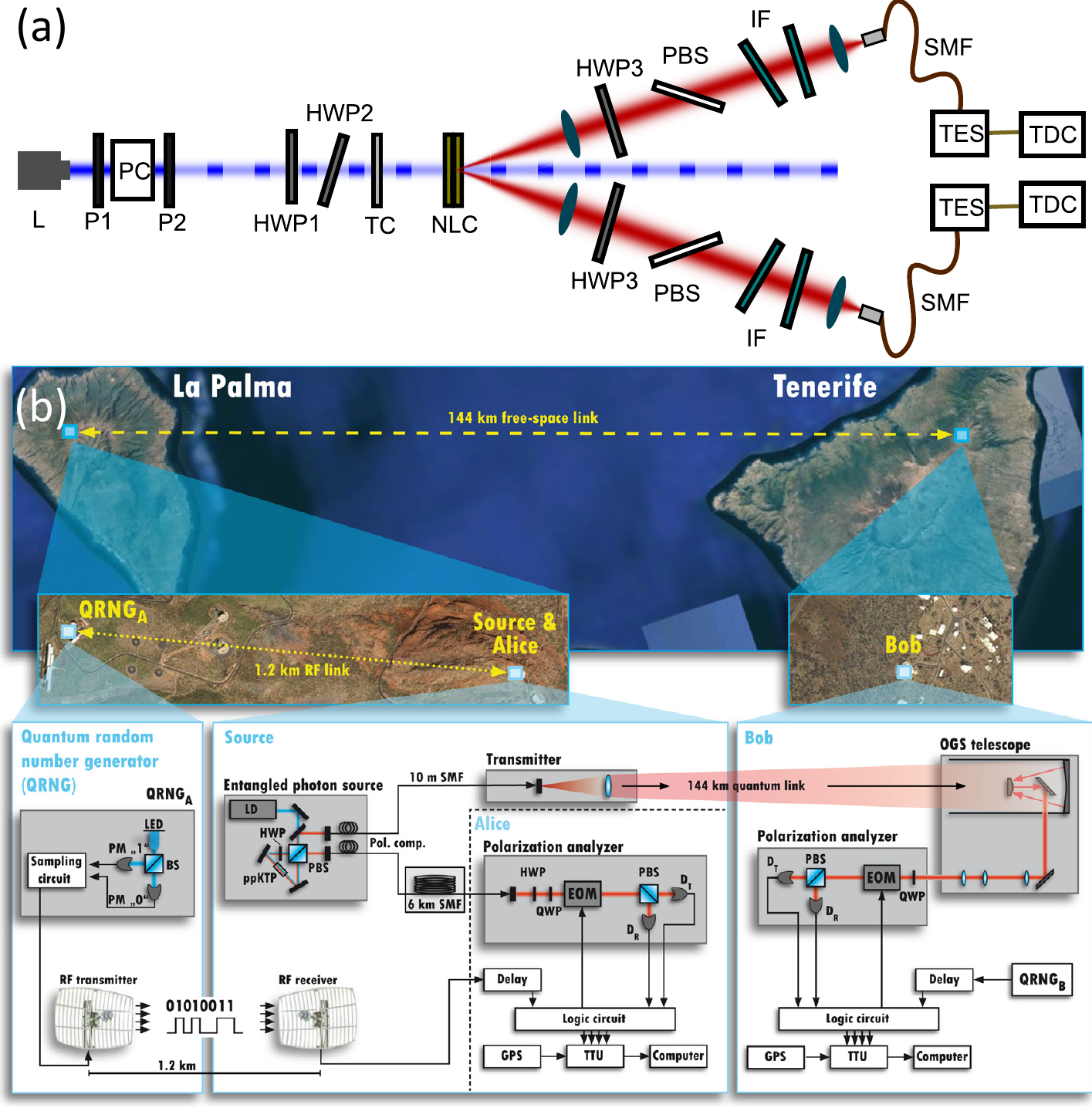}
\caption{{\textbf{Experiments for closing loopholes.}
(a) High detection efficiency can \blueit{be} achieved with TES to close the detection loophole. Figure from Ref.~\onlinecite{ch-prl-111-130406}.
(a) Space-like separating the quantum random number generators that \blueit{choose} the random measurement settings and the measurement apparatus, enables the locality and freedom of choice experiment loopholes to \blueit{be} closed. Figure from Ref.~\onlinecite{sc-pnas-107-19708}.
}}
\label{SetupFig2}
\end{figure}

Two \blueit{recent} experiments~\cite{gi-nat-497-227,ch-prl-111-130406} report violation of Eberhard's inequality to close the detection loophole. Both experiments use high efficiency single photon detecting transition edge sensors (TES) \cite{li-oe-16-3032} and high collection efficiency photon sources to surpass Eberhards efficiency threshold, and each obtain $\eta > 70 \%$: Ref.~\onlinecite{gi-nat-497-227} uses a high collection efficiency photon source based on a Sagnac configuration {\cite{ki-pra-7312316,fe-oe-15-15377}; Ref.~\onlinecite{ch-prl-111-130406} uses a non-colinear SPDC photon source configuration (Fig.~\ref{SetupFig2} (a)). While both demonstrations are of sufficient efficiency to close the detection loophole,
the experiment in Ref.~\onlinecite{gi-nat-497-227} is still open to the so-called \textit{``coincidence-time''} \blueit{loophole \cite{la-epl-67-707}}. This experiment (and numerous others) rely on using timing windows defined by single photon detection events to perform coincident detection analysis: when one photon detector registers a single-photon event, a coincidence event is recorded if a second single-photon event is recorded within a prescribed window of time.  \blueit{The coincidence time loophole allows the detection time to be shifted by the local measurement settings in or out of the coincidence window, so that a completely local process can match quantum mechanical expectation values.  However, } this loophole can be avoided by using a coincidence window defined around a system clock: Ref.~\onlinecite{ch-prl-111-130406} achieves this with a chopped laser pulse that drives the SPDC to create photon pairs in well-defined events.

Ref. \onlinecite{ch-prl-111-130406} also \blueit{highlights} the \textit{``production-rate loophole''} where non-random drift\blueit{ing of the} pump laser power or detection efficiency can be exploited by local realistic models, however the experimental drifts in Ref.~\onlinecite{gi-nat-497-227} have been shown~\cite{ko-arxiv:1307.6475} to be not sufficient \blueit{for this loophole}. Alternatively, a quantum random number generator \blueit{ can} be used to randomly choose measurement settings in order to close the production-rate loophole~\cite{ch-prl-111-130406}. Furthermore, satisfying the more stringent requirement of randomly chosen measurement settings \textit{for every} entangled particle pair in order to close the \textit{freedom of choice loophole} simultaneously addresses production-rate loophole.

\emph{Freedom of choice and locality loophole---}Two famous experiments attempted to close the \textit{locality loophole} through space like separation by fast measurement settings chosen during the time of flight of the entangled photons \cite{as-prl-49-1804, we-prl-81-5039}. However the settings of Ref.~\onlinecite{as-prl-49-1804} were chosen using periodic sinusoid\blueit{s} and were therefore predictable and susceptible to influence by hidden variables created at the source, \blueit{so} failed to close the \textit{freedom of choice loophole}---the possible influence of measurement settings by either other measurement apparatus or hidden variables created at the source of photons. The random settings of \blueit{Ref.~\onlinecite{we-prl-81-5039}} were chosen within the forwards light cone of the emission point of the entangled photons, \blueit{so could also} have been influenced by hidden variables created at the source. 
Improving upon these experiments the authors of Ref.~\onlinecite{sc-pnas-107-19708} space-like separate their random number generators to remove the possibility for transmitting any physical signal between entangled particle emission and the random \blueit{measurement settings}.
\blueit{T}his \blueit{Bell test was performed} between the two Canary Islands La Palma and Tenerife separated by 144km, with the quantum random number generator used to choose measurement bases space-like separated from the rest of the experiment {(Fig.~\ref{SetupFig2} (b))}.

\emph{EPR-steering---}Almost 80 years after Schr\"{o}dinger referred to the effects of entanglement as ``piloting'' or ``steering'' of one quantum state by the measurement of another, the concept of EPR-steering was formalised \cite{wi-prl-98-140402,jo-pra-76-052116} and followed swiftly by an EPR-steering inequality based on local models \cite{ca-pra-80-032112}. Steering sits strictly between entanglement witnesses~\cite{pl-qic-7-1}---that rely entirely on assumptions that quantum mechanics is correct, to test for the presence of non-seperability---and Bell-nonlocality---ideally no assumptions are made about the experimental setup or the model of physics. The modern concept (Fig.~\ref{SteeringFig}) assumes that one half of the system, an observer Bob, fully trusts his measurement apparatus and \blueit{that} any states in his possession adhere to the laws of quantum mechanics. A second party (Alice) is tasked with convincing Bob that she can steer a quantum state that she has already sent to him. Importantly, no assumptions are made about the physics to which Alice has access, so she \blueit{is} free to use any means to carry out her task. 
Assuming local models, this experiment is constrained by the inequality~\cite{ca-pra-80-032112}
\begin{eqnarray}
S_n \equiv \frac{1}{n}\sum_{k=1}^n \left\langle A_k \sigma_k^B\right\rangle \leq C_n
\label{SteeringInequality}
\end{eqnarray}
in which Alice and Bob compare $n$ measurement results; $\sigma_k^B$ is the $k^\textrm{th}$ of $n$ measurements performed by Bob in conjunction with Alice declaring a measurement result $A_k\in\{-1, +1\}$.
$C_n$ is the maximum value that can be obtained for the quantity $S_n$, provided Bob \blueit{has} pre-existing state\blueit{s} known to Alice. This inequality is violated when Alice instead shares entanglement with Bob, and through her own measurements, affect\blueit{s} \blueit{Bob's} results.

Saunders \textit{et al.}~\cite{sa-nphys-6-845} performed the first experimental demonstrations of violating Steering inequalities with polarisation entangled photons, showing that increasing the number of measurements $n$ (testing up to $n=6$) increases the robustness of this nonlocality test to experimental noise. However, just like Bell-like inequalities, local models can also exploit loopholes to explain steering. Steering has less stringent requirements than the aforementioned nonlocality tests, due to the asymmetry of the experiment, and has \blueit{an} experiment efficiency threshold of $\eta>1/3$ for closing the detection loophole when using $n=3$ measurements. Three \blueit{experiments published around the same time}
collectively address loop-hole-free steering~\cite{be-prx-2-031003,sm-ncomm-3-625,wi-njp-14-053030}.
All three experiments use Sagnac entanglement sources (see Box 1) to increase experiment efficiency and close the detection loophole; in addition \mbox{Smith \textit{et al.}~\cite{sm-ncomm-3-625}} use TES single photon detectors. 
\mbox{Bennet \textit{et al.}~\cite{be-prx-2-031003}} use up to $n=16$ measurement settings, which they show allows them to measure violation of Eq.~(\ref{SteeringInequality}) without assuming fair sampling despite high loss (87\% loss) induced by 1km of coiled optical fibre between the entanglement source and one of the measurement apparatus; this explores the conditions for closing freedom of choice and locality loopholes over a lossy channel.
In addition to closing the detection loophole, 
Witmann~\cite{wi-njp-14-053030} enforces strict Einstein locality conditions, with space-like separation over 48m of optical fibre. This closes the locality loophole. In addition they use a space-like separated quantum-random number generator,  closing the freedom-of-choice loophole that would otherwise allow the photon pair source to influence the choice of measurement setting.
This is the first time a nonlocal quantum effect has been explored whilst simultaneously closing three major loopholes. 
Collectively, these experiments~\cite{be-prx-2-031003,sm-ncomm-3-625,wi-njp-14-053030} mark an important progression towards loophole-free device-independent tests \blueit{of} Bell nonlocality.

\begin{figure}[t!]
\centering
\includegraphics[width=\columnwidth]{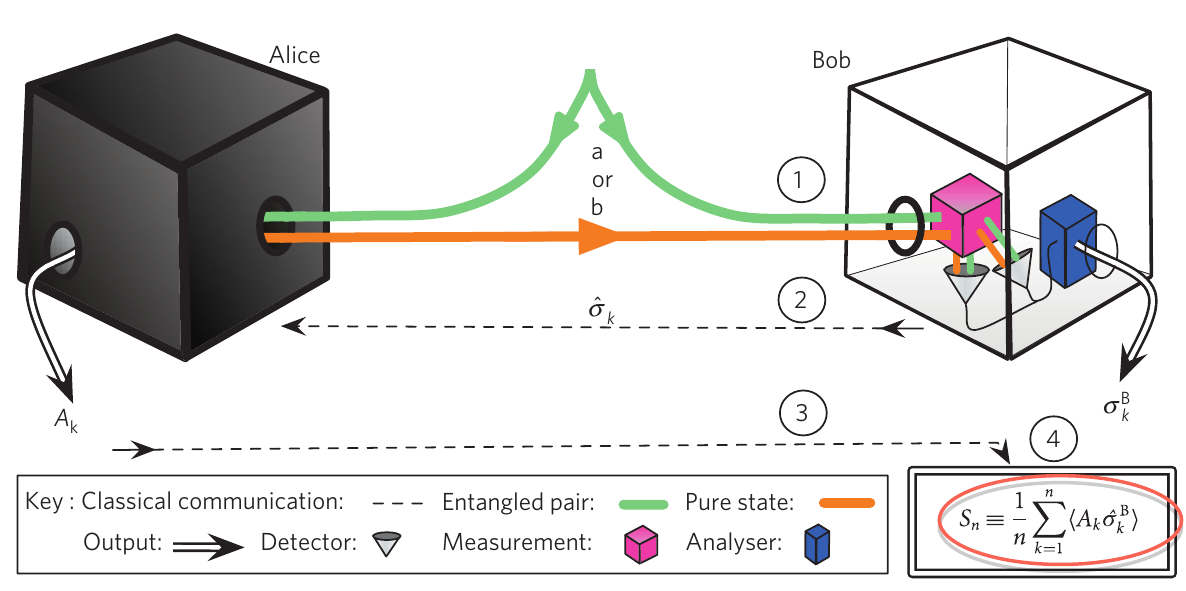}
\caption{\textbf{EPR steering.} Here one observer (Bob) trusts their system works according to quantum mechanics (denoted by a clear box), while another party (Alice) is tasked with supplying Bob with a quantum state and demonstrating \blueit{she} can affect his measurement results by any means (black box). Assuming local laws of physics, \blueit{an} inequality for this experiment \blueit{is derived}, which is violated when Alice chooses to share entanglement between herself and Bob. Figure from Ref.~\onlinecite{sa-nphys-6-845}.}
\label{SteeringFig}
\end{figure}

\emph{Reference frame independent nonlocality tests---}Traditionally, nonlocality tests take place within a shared reference frame. That is to say that Alice and Bob are able to align their measurement apparatus with respect to one another.
This may be problematic for experiments using long optical fibre or free-space orbital communications.  One solution is to harness decoherence-free subspaces \cite{ca-prl-91-230403}, which has been recently implemented using hybrid polarisation and orbital angular momentum entangled states to violate a Bell inequality in alignment free settings \cite{da-natcomm-3-961}. This requires an increase in dimension of the quantum system investigated. Remarkably, it has been shown that sharing a complete reference frame is not required for two remote parties to violate a Bell inequality, provided the parties share one measurement direction perfectly, they have high probability to violate a Bell inequality perfectly with a maximally entangled state by each choosing maximally complimentary measurements in the plane orthogonal to the shared direction in the Bloch sphere \cite{pa-pra-86-032322}. By increasing the complexity of the measurements each party makes, observers can always violate a Bell inequality without sharing any information about their reference frames \cite{sh-scirep-2-470, wa-pra-85-024101}. This potentially removes the need of establishing reference frames for future nonlocality tests, in particular taking nonlocality tests into orbit to help address the locality loophole.

\emph{Multipartite locality tests---}The majority of nonlocality tests have been focused towards using bi-partite entanglement. Greenberger Horne and Zeilinger extended nonlocality tests to that of three-party entanglement (GHZ); this was formulated into an inequality to test for multipartite nonlocality by Mermin~\cite{me-prl-65-1838}.  The first 3-photon GHZ entanglement was demonstrated fifteen years ago using a pulsed SPDC source~\cite{bo-prl-82-1345} and was then subsequently used to violate Mermin's inequality~\cite{pa-na-403-515}; four-photon GHZ states~\cite{pa-prl-86-4435,ei-prl-90-200403} have also been used for local realism tests~\cite{zh-prl-91-180401}. However, until recently no multi-photon experiment has succeeded in addressing loopholes that can be exploited by LHV. The major contributing factor is the typically low brightness of multi-photon entangled sources.
Recently, \mbox{Erven \textit{et al.}~\cite{er-arXiv:1309.1379v1}} reported generating heralded three-photon Greenberger-Horne-Zeilinger entanglement at sufficient rates ($40$ Hz) to distribute the three photons using optical fibre and free-space links to independent measurement stations to violate Mermin's inequality. With sufficiently separated measurement stations and entanglement source, the authors address the locality loophole while the freedom of choice loophole is closed by spatially separating a random number generator that defines the measurement basis settings. However, \blueit{experiment} efficiencies below the threshold required to close the loophole of Mermin's inequality mean the detection loophole is not closed and fair-sampling is assumed. This leaves open the possibility of employing high efficiency photon detectors and developing efficient collection in multi-photon entangled \blueit{states} for loophole-free multi-partite nonlocality tests in the future.

\noindent\textbf{Box 1: Enabling technology for current and future nonlocality tests}

\emph{Sources of entangled photons---}For nearly two decades, spontaneous parametric downconversion based \blueit{on nonlinear crystals}
has been the most \blueit{widely} applied source of entanglement in quantum optics \cite{kw-prl-75-4337}. A configuration that has recently advanced collection efficiency is \blueit{that of} collinear SPDC, pumped in both directions coherently with a laser split at a beamsplitter, in a polarisation \blueit{S}agnac \blueit{interferometer}~\cite{ki-pra-7312316,fe-oe-15-15377} (Fig.~\ref{Tech}). This allows inherent stability without the need for active stabilisation. By eliminating the transverse walk-off effect via periodically poling in the nonlinear crystal, high collection efficiency into single mode fibre is obtained. This configuration has been demonstrated to operate with CW or pulsed regimes \cite{pr-oe-20-25022} with at least 80\% coupling efficiency and for a number of nonlinear materials \cite{gu-arxiv-1309-2457}. For future experiments, a more compact \blueit{source} of entangled photons would \blueit{likely} use an integrated architecture, where stabilised path-entangled photons~\cite{si-nphoton-aop} and polarisation-entangled photons \cite{ma-srep-2-817} can be generated. 

\emph{Increasing detector efficiency---}Single photon detectors~\cite{ha-nphoton-3-696} underpin the measurement\blueit{s} \blueit{made by} the observers in any photonic nonlocality experiment. 
Transition edge sensors (TES) are fabricated using a thin tungsten film embedded in an optical stack of materials to enhance the absorption \cite{li-oe-16-3032}. With the voltage biased at their super-conducting transition, absorbed photons cause a measurable change in the current flowing through the tungsten film that is efficiently measured with a superconducting quantum interference device amplifier. TES require cooling to $\sim 100$ mK using adiabatic demagnetization refrigerators and detection efficiencies of $\sim95\%$ are now routinely reported. Nanowire superconducting single-photon detectors (SSPDs)\cite{go-apl-79-705} have emerged as a promising alternative for both free-space and integrated applications: here a single photon absorbed by a superconductor biased just below its critical current $I_c$ creates a local resistive ``hotspot'', generating a voltage pulse. Superconducting detectors based on NbN nanowires operate at $\sim4$K temperatures and are capable of very fast counting rates (up to GHz) \blueit{and} low dark counts ($<1$ Hz)~\cite{go-apl-79-705}. NbN nanowire detectors can operate in commercial cryo-coolers \cite{ha-nphoton-3-696}. Recent NbN nanowire detectors using a travelling wave design \cite{pe-ncomm-3-1325} have demonstrated on-chip detection efficiency above 90\%. In addition, recent realisation of NbTiN nanowire single photon detectors on SiN \cite{sc-scirep-2-1893} extends the operating wavelength from IR to visible and reduces dark count rate to milli-Hz.

\begin{figure}[t!]
\centering
\includegraphics[width=\columnwidth]{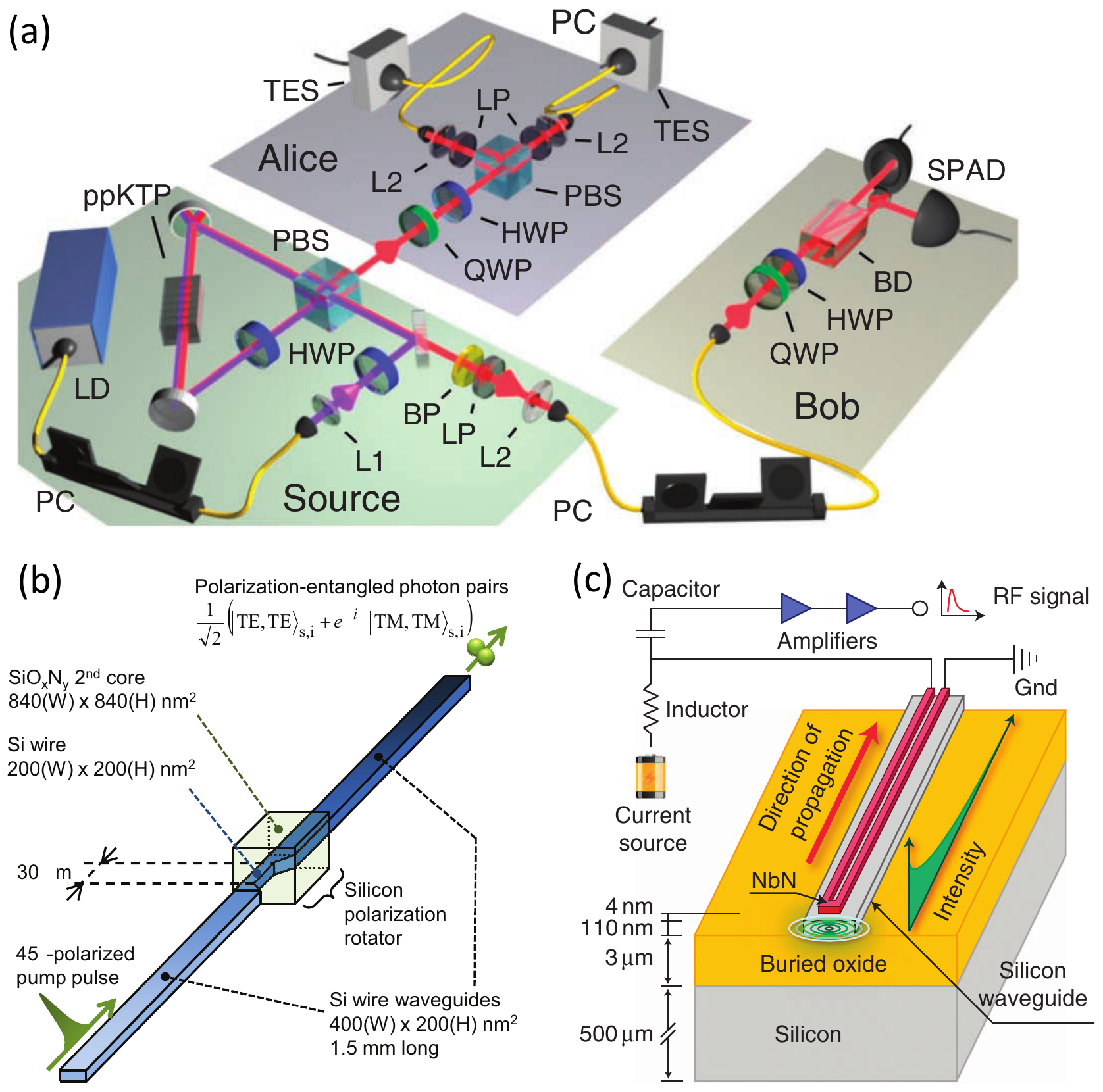}
\caption{\emph{\textbf{Box figure:}} \textbf{Current and future photonics for nonlocality tests.} (a) A  polarisation Sagnac interferometer configuration can enable collection efficiency of generated photon pairs into optical fibre with $>80\%$, which when used together with high efficiency $> 95\%$ TES single photon detectors can be used to address the detection loophole in nonlocality tests. The example here depicts Alice using TES, to address the detection loophole in EPR steering in the experiment reported by Ref.~\onlinecite{sm-ncomm-3-625}.  (b) Waveguide entangled sources offer potentially repeatable, high brightness and high efficiency sources of entanglement. Figure from Ref.~\onlinecite{ma-srep-2-817}. (c) SSPDs offer a 4K alternative in high efficiency and fast single photon detection that can be monolithically integrated into waveguide \blueit{structures} for potentially compact photonics measurement apparatus. Figure from Ref.~\onlinecite{pe-ncomm-3-1325}. }
\label{Tech}
\end{figure}

\clearpage
\newpage

\noindent
\textbf{Outlook}\\
Photonic experiments over the past four decades have answered many important debates in the fundamental theory of quantum mechanics and new photonic technologies continue to create opportunities to close loopholes, answer old questions, and even inspire new theoretical research.  Experimental confirmation of the predictions of quantum physics during the previous century forced a reevaluation of the understanding of the operation of the universe as a classical machine, at least at the microscopic scale.  Over the coming decades, as we increase our capabilities to harness the effects of quantum mechanics to build quantum computers \cite{Nielsen:2011vx}, we will test the extent to which quantum effects persist at a macroscopic scale, with further potential consequences for our understanding of the universe.  Famously, the Extended Church Turing Thesis (ECT) says that all computational problems that are efficiently solvable with realistic physical systems can be efficiently solved with a classical machine --- a statement clearly in conflict with our hopes for the capabilities of quantum computers \cite{Shor:1994ul}.  While we might have to wait some time for a universal quantum computer to operate at the scale which challenges the ECT, recent theoretical \cite{Aaronson:2011tj} and technological advances in quantum photonics \cite{politi2008} have developed a path to challenging the ECT on a near term timescale, with a non-universal quantum photonic device that performs a task known as Boson Sampling \cite{Broome:2013ti, Spring:2013to,Crespi:2012fu,Tillmann:2012ux}.  If experiments confirm the prediction, as we believe they will, that our universe cannot be efficiently simulated by a classical machine, then there may be other confounding features of quantum mechanics currently hidden from us and apparent only through simulations on a quantum computer.  It is therefore possible that, rather than confirming existing theory, future photonic experiments might be the first to reveal new and complex quantum phenomena, requiring innovative theoretical explanations.

\begin{acknowledgements}
The authors are grateful for financial support from EPSRC, ERC, NSQI, NRF (SG) and MOE (SG). JCFM is supported by a Leverhulme Trust Early-Career Fellowship. JLOB acknowledges a Royal Society Wolfson Merit Award and a Royal Academy of Engineering Chair in Emerging Technologies. 
\end{acknowledgements}

\end{document}